# Magnetoelectric effects and valley controlled spin quantum gates in transition metal dichalcogenide bilayers

Zhirui Gong[1,2*], Gui-Bin Liu[1,2,3*], Hongyi Yu[1,2], Di Xiao[4], Xiaodong Cui[1], Xiaodong Xu[5,6], and Wang Yao[1,2†]

[1] Department of Physics, The University of Hong Kong, Hong Kong, China

[2] Center of Theoretical and Computational Physics, The University of Hong Kong, Hong Kong, China

[3] School of Physics, Beijing Institute of Technology, Beijing 100081, China

[4] Department of Physics, Carnegie Mellon University, Pittsburgh, Pennsylvania, USA

[5] Department of Physics, University of Washington, Seattle, Washington, USA

[6] Department of Material Science and Engineering, University of Washington, Seattle, Washington, USA

* These authors contributed equally to the work.

† Email: wangyao@hku.hk

**Abstract:** In monolayer group-VI transition metal dichalcogenides (TMDC), charge carriers have spin and valley degrees of freedom, both associated with magnetic moments. On the other hand, the layer degree of freedom in multilayers is associated with electrical polarization. Here, we show that TMDC bilayers offer an unprecedented platform to realize a strong coupling between the spin, layer pseudospin, and valley degrees of freedom of holes. Such coupling not only gives rise to the spin Hall effect and spin circular dichroism in inversion symmetric bilayer, but also leads to a variety of magnetoelectric effects permitting quantum manipulation of these electronic degrees of freedom. Oscillating electric and magnetic fields can both drive the hole spin resonance where the two fields have valley-dependent interference, making possible a prototype interplay between the spin and valley as information carriers for potential valley-spintronic applications. We show how to realize quantum gates on the spin qubit controlled by the valley bit.

**Introduction**

Device applications exploiting internal degrees of freedom of charge carriers may lead to new electronic technologies outperforming their conventional counterparts which rely on charge flow only. A seminal example is electron spin, which has been extensively exploited as a carrier of classical or quantum information[1-3]. Recently, the emergence of atomically thin two-dimensional (2D) crystals such as graphene and transition metal dichalcogenides (TMDC) have offered new playgrounds to explore novel electronic device concepts[4-10]. In addition to spin, two types of internal indices of electrons have been investigated as information carriers, namely the layer[11-14] and valley[15-19] degrees of freedom.

The *Layer* degree of freedom exists in bilayer systems and can be described as a pseudospin. Pseudospin up (down) refers to the state where the charge carrier is located in the upper (lower) layer. Thus, pseudospin polarization directly corresponds to electrical polarization. Such layer pesudospin has been proposed in graphene bilayers for pseudospintronics[11-14]. *Valley* refers to the degenerate extrema of energy bands, which in many hexagonal 2D crystals are located at the corners of the hexagonal Brillouin zone (*K* points). It was predicted that in the absence of inversion symmetry valley degree of freedom can be associated with magnetic moment and optical circular dichroism[18,20], which has made possible the first observations of optical pumping of valley polarization in TMDC monolayers[21-23] and biased bilayers[24]. Since the spin and valley have magnetic moment which can be controlled by magnetic and optical means, while the layer pseudospin corresponds to an electrical polarization subject to electrical manipulation, a system allowing interplay of these different degrees of freedom may offer unprecedented possibilities to exploit their quantum control for new device concepts.

In this letter, we show that in group-VI TMDC bilayers, the spin, the layer pseudospin, and the valley degrees of freedom are strongly coupled for the holes in the *K* valleys, leading to a variety of magnetoelectric effects to realize valley-spintronics. Such a coupling suppresses the interlayer hopping, making possible the spin Hall effect and spin circular dichroism even in the presence of inversion symmetry. A static magnetic field can induce oscillations of the layer (electrical) polarization, while spin precession in magnetic field can be electrically controlled. The frequencies of the spin and layer dynamics become valley dependent when both the magnetic and electric fields are present, giving rise to beating phenomena where a spin polarization can evolve from zero net polarization. An oscillating electric field and magnetic field will both drive the hole spin resonance. Interestingly, the two fields can have a valley-dependent interference through their coupling to the same transition. Such a phenomenon not only achieves an effective coupling between the oscillating electric and magnetic fields, but also makes possible a prototypical interplay between the spin and valley information carriers where quantum gates on the spin qubit are controlled by the valley bit.

**Results**

**Strong coupling between spin, valley, and layer pseudospin.** Group-VI TMDC bilayers are AB stacked: one monolayer sits on another but with 180° rotation (Fig. 1(a)). Pristine bilayers are therefore inversion symmetric. As in monolayers, *ab initio* calculations show that the valence and conduction band edges near $K$ points in bilayers are dominantly contributed by $d_{z^2}$, $d_{xy}$, and $d_{x^2-y^2}$ orbitals of metal atoms. A minimal band model of bilayers in the neighborhood of $K$ points can be constructed by adding interlayer hopping to the $k \cdot p$ model of monolayers established in Ref. [20],

$$H(q) = \begin{bmatrix} \Delta & at(\tau_z q_x + iq_y) & 0 & 0 \\ at(\tau_z q_x - iq_y) & -\tau_z s_z \lambda & 0 & t_\perp \\ 0 & 0 & \Delta & at(\tau_z q_x - iq_y) \\ 0 & t_\perp & at(\tau_z q_x + iq_y) & \tau_z s_z \lambda \end{bmatrix}. \quad (1)$$

The basis is $\{|d_{z^2}^u\rangle, 1/\sqrt{2}(|d_{x^2-y^2}^u\rangle - i\tau_z|d_{xy}^u\rangle), |d_{z^2}^l\rangle, 1/\sqrt{2}(|d_{x^2-y^2}^l\rangle + i\tau_z|d_{xy}^l\rangle)\}$, where the superscripts "$u$" and "$l$" denote the "upper" and "lower" layer respectively. $q$ is the relative wavevector with respect to the $K$ points, $\Delta$ the monolayer band gap, $a$ the lattice constant, $t$ the nearest-neighbor intra-layer hopping, with $\lambda$ the spin-valley coupling of holes in monolayers. $t_\perp$ is the interlayer hopping for holes, while the interlayer hopping for electrons vanishes at $K$ points due to the symmetry of the $d_{z^2}$ orbital. $\tau_z = \pm 1$ is the valley index of bilayer bands and $s_\mu$ denotes the Pauli matrices for the spin. Similar to monolayers[20], the spin up and spin down states are still decoupled in bilayers since interlayer hopping conserves spin. Table 1 lists these parameters obtained by fitting the *ab initio* band structures.

The Hamiltonian of holes near $K$ points can be obtained through a canonical transformation of Eq. (1) eliminating the interband coupling,

$$H_v(q) = -\varepsilon_q + \lambda\left(-1 + \frac{\varepsilon_q}{\Delta}\right)\tau_z\sigma_z s_z + t_\perp\left(1 - \frac{\varepsilon_q}{\Delta}\frac{\Delta^2+\lambda^2}{\Delta^2-\lambda^2}\right)\sigma_x, \tag{2}$$

where $\varepsilon_q = \Delta\frac{a^2 t^2 q^2}{\Delta^2-\lambda^2}$ gives the energy dispersion. The small quantity $\varepsilon_q/\Delta \ll 1$ will be neglected hereafter. $\sigma_\mu$ are Pauli matrices defined in the basis $\left\{\frac{1}{\sqrt{2}}\left(|d^u_{x^2-y^2}\rangle - i\tau_z|d^u_{xy}\rangle\right), \frac{1}{\sqrt{2}}\left(|d^l_{x^2-y^2}\rangle + i\tau_z|d^l_{xy}\rangle\right)\right\}$, which can be regarded as the *layer pseudospin*. The second term in Eq. (2) represents a strong coupling between the layer pseudospin ($\sigma_z$), the real spin ($s_z$) and the valley ($\tau_z$), while the third term is the interlayer hopping. The coupling between the three indices is the manifestation of the monolayer spin-valley coupling in the AB stacking order. This coupling is much stronger than (comparable with) the interlayer hopping in $WX_2$ ($MoX_2$) as listed in Table I, and it originates from the strong spin-orbit interaction in the $d$-orbitals of the metal atoms. A direct consequence is that the interlayer hopping is virtually suppressed as observed in $WX_2$ multilayers[25].

At each wavevector $q$, the eigenstates of Eq. (2) are two spin doublets separated by a large energy $\sqrt{\lambda^2 + t_\perp^2}$. We focus on the spin doublet at the band edge (enclosed by the dashed box in Fig. 1(c)), which in valley $K$ is given by

$$|K,\uparrow\rangle = \frac{1}{\sqrt{2}}\left(\sin\alpha\left(|d^u_{x^2-y^2}\rangle - i|d^u_{xy}\rangle\right) + \cos\alpha\left(|d^l_{x^2-y^2}\rangle + i|d^l_{xy}\rangle\right)\right)\otimes|\uparrow\rangle,$$

$$|K,\downarrow\rangle = \frac{1}{\sqrt{2}}\left(\cos\alpha\left(|d^u_{x^2-y^2}\rangle - i|d^u_{xy}\rangle\right) + \sin\alpha\left(|d^l_{x^2-y^2}\rangle + i|d^l_{xy}\rangle\right)\right)\otimes|\downarrow\rangle. \tag{3}$$

$|\uparrow\rangle$ and $|\downarrow\rangle$ denote the spin up and down states respectively. The spin doublet $\{|\bar{K},\downarrow\rangle, |\bar{K},\uparrow\rangle\}$ in valley $-K$ is the time reversal of the above. These energy eigenstates are associated with a spin and valley dependent layer (electrical) polarization,

$$\langle\sigma_z\rangle = -s_z\tau_z\cos 2\alpha, \quad \cos 2\alpha \equiv \frac{\lambda}{\sqrt{\lambda^2+t_\perp^2}}, \tag{4}$$

Table 1 lists the computed values of $\cos 2\alpha$ using the parameters $t_\perp$ and $\lambda$ fitted from the *ab initio* band dispersion, which agrees well with the layer polarization directly evaluated from the *ab initio* wavefunction. This shows that our minimal model gives a good description of the electronic states near $K$ points. The large layer polarization means that the spin doublet is predominantly localized in (opposite) individual layers, and interlayer hopping of holes is largely suppressed.

An important consequence of the spin dependent layer polarization is that bilayers will inherit most of the spin physics of monolayers. In Fig. 1(c), we illustrate the optical interband transitions by circularly polarized light between the valence and conduction band edges near $K$ points. Spin is conserved here by the optical transition. In both valleys, $\sigma+$ ($\sigma-$) polarized light predominantly excites spin up (down) photocarriers[22]. We find the degree of spin polarization created upon absorption of a circularly polarized photon to be $\cos 2\alpha$, directly corresponding to the layer polarization of the hole states. This spin circular dichroism can be used for optical injection and detection of spin polarization near $K$ points. The valley circular dichroism is absent in pristine bilayers because of inversion symmetry. Moreover, the Berry curvature of the band edge holes at $K$ points is:

$$\Omega = s_z \frac{2a^2 t^2}{\left(\Delta - \sqrt{\lambda^2 + t_\perp^2}\right)^2} \cos 2\alpha, \qquad (5)$$

which depends only on the spin index. Thus the spin Hall effect of holes is also present in inversion symmetric bilayers, while the valley Hall effect vanishes.

**Magnetoelectric effects.** While the spin couples to real space magnetic fields, the layer (electrical) polarization couples to electric fields in the perpendicular ($z$) direction. The coupling between spin and layer pseudospin therefore gives rise to magnetoelectric effects. The hole Hamiltonian in the presence of electromagnetic fields is,

$$H_v = -\lambda \tau_z \sigma_z s_z + t_\perp \sigma_x + B_x s_x + B_z s_z + E_z \sigma_z. \qquad (6)$$

Without loss of generality, the in-plane component of the magnetic field is taken to be along the $x$-direction. $B_x$, $B_z$ and $E_z$ are strengths of magnetic and electric fields in units of energy, normalized by the magnetic and electric dipole respectively. $B_x$ and $B_z$ can be up to ~ meV for magnetic fields up to a few Tesla. A moderate electric field is considered where $E_z$ has comparable magnitude to $B_x$ and $B_z$, so $\lambda, t_\perp \gg B_x, B_z, E_z$. We focus on the spin doublet at the band edge (c.f. Eq. (3)), and the Hamiltonian in Eq. (6) projected into the subspace of this doublet reads,

$$H_D = \begin{bmatrix} -\tau_z E_z \cos 2\alpha + B_z & B_x \sin 2\alpha \\ B_x \sin 2\alpha & \tau_z E_z \cos 2\alpha - B_z \end{bmatrix}. \quad (7)$$

where the basis is $\{|K \uparrow\rangle, |K \downarrow\rangle\}$ at valley $K$ ($\tau_z = 1$) and $\{|\overline{K} \uparrow\rangle, |\overline{K} \downarrow\rangle\}$ at valley $-K$ ($\tau_z = -1$). The effective fields are illustrated in Fig. 2(a) and 2(d) for valley $K$ and $-K$, respectively.

As the doublet is associated with spin and valley dependent layer (electrical) polarization, its coupling to both the electric and magnetic fields gives rise to various forms of magnetoelectric effects. A few examples are given. Fig. 3(a) shows that spin precessions in a magnetic field can be controlled by the electric field $E_z$. Here the initial state is a spin polarized one with no layer and valley polarizations: $\rho_S = 0.5|K \uparrow\rangle\langle K \uparrow| + 0.5|\overline{K} \uparrow\rangle\langle \overline{K} \uparrow|$. Fig. 3(b) shows that static magnetic field with finite in-plane component can induce oscillations of the electrical polarization $\langle \sigma_z \rangle$. The initial state is a layer polarized one with no spin and valley polarizations: $\rho_E = 0.5|K \uparrow\rangle\langle K \uparrow| + 0.5|\overline{K} \downarrow\rangle\langle \overline{K} \downarrow|$. We note that spin polarization can be prepared by optical pumping with circularly polarized light (cf. Fig. 1(c)), while layer polarization can be prepared by tunneling from the lower side of the bilayer as interlayer hopping is virtually suppressed.

When both $E_z$ and $B_z$ are finite, the direction and the magnitude of the total effective field become valley dependent (c.f. Eq. (7)), resulting in different oscillation frequencies in the two valleys as illustrated in Fig. 2. This gives rise to beating phenomena in the oscillation of spin and electrical polarizations when both valleys are populated, as shown in Fig. 3. Remarkably, because of the beating, a finite spin polarization can evolve out of an initial state with zero spin polarization in all directions (see Fig. 3(c)).

We note that an arbitrary linear superposition of the doublet is in fact an entangled state between the spin and the layer pseudospin as shown in Eq. (3). After tracing out the layer degrees of freedom, the magnitude of the in-plane spin component will scale down by a factor of $\sin 2\alpha$, and the spin vector always lies within an elliptical sphere (c.f. Fig. 2(b) and 2(e)).

**Valley controlled quantum gate of spin qubit**. When static electric field $E_z$ or magnetic field $B_z$ is applied to split the doublet along the $z$-axis, an oscillating in-plane magnetic field $B_x$ can drive the hole spin resonance. The doublet splitting becomes valley dependent when both $E_z$ and $B_z$ are finite (c.f. Eq. (7) and Fig. 4(a)). Thus the hole spin resonance can be selectively addressed at valley $K$ or $-K$ by choosing the frequency of $B_x$.

As a quantum two-level system, the doublet can be used as a spin qubit for information processing. The valley index of the doublet, on the other hand, represents another bit of information. A prototype interplay between the two types of information carriers for potential valley-spintronic applications can be made possible. For example, as shown in Fig. 4(a), we consider the application of a pulsed $B_x$ with central frequency on resonance with the splitting in valley $-K$. If the valley bit is in state $-K$, the spin qubit is coherently rotated about the $x$-axis by the pulse. If the valley bit is in state $K$, the spin qubit is unchanged because of the large detuning. This valley dependent rotation represents a quantum gate on the spin qubit controlled by the valley bit. If the rotation is $\pi$, a valley controlled NOT gate on the spin qubit is realized. We note that this controlled NOT gate can be used for deterministic conversion between an electrical polarization and a spin polarization (e.g. between $\rho_E$ and $\rho_s$).

**Interference between oscillating electric and magnetic fields.** If a static $B_x$ is applied to split the doublet along the $x$-axis, both the oscillating electric field and magnetic field in the $z$-direction can drive the hole spin resonance[26-30]. Pulsed oscillating $E_z$ or $B_z$ can therefore realize coherent rotations of the spin qubit about the $z$-axis. Interestingly, interference between the oscillating electric field and magnetic field can now be realized through their coupling to the same transition. The interference is valley dependent as the coupling of the spin qubit to $E_z$ has a valley-dependent sign. If $E_z(t)$ and $B_z(t)$ are in phase with the same frequency, their interference is destructive in valley $K$ and constructive in valley $-K$. Such a phenomenon induces an effective coupling between the oscillating electric field and magnetic field in the presence of valley polarization. The valley-dependent interference can also be utilized for valley-controlled spin qubit rotations about the $z$-axis, as shown in Fig. 4(b).

**Discussions**

The phenomena above are predicted for holes in the $K$ valleys. Holes can be optically injected into $K$ valleys in group-VI TMDC bilayers by pumping the direct transition between the higher lying valence band and the conduction band. The strong PL peak associated with this transition suggests high efficiency of the injection[21,22,24,25]. Interestingly, *ab initio* calculations show that although WSe$_2$ bilayer has an indirect gap, the valence band maxima are at the $K$ points (see supplementary information). For MoSe$_2$ and WS$_2$ bilayers, hole pockets are also expected to appear at $K$ points at low doping. Thus lightly $p$-doped WSe$_2$, MoSe$_2$ and WS$_2$ bilayers could be ideal platforms to explore these phenomena.

Long lifetimes of the polarizations are also essential for valley-spintronic applications. Spin and valley relaxations in TMDC bilayers can be qualitatively different from those in monolayers[31], and remain open questions. Notably, for holes near $K$ points, spin flip ($|K \uparrow\rangle \leftrightarrow |K \downarrow\rangle$) and valley flip ($|K \uparrow\rangle \leftrightarrow |\bar{K} \uparrow\rangle$) are suppressed by the spin and valley dependent layer polarization from the coupling of the three indices. Since $|K \downarrow\rangle$ and $|\bar{K} \uparrow\rangle$ are located on the layer opposite to $|K \uparrow\rangle$, the spatial overlap of the initial and final state wavefunctions of spin and valley flips is small, which is given by $\sin^2 2\alpha \sim 0.1$ ($\sim 0.3$) in WX$_2$ (MoX$_2$) bilayers. This small factor helps to reduce both spin and the valley relaxations. Moreover, long valley lifetime has also been suggested by the observation of valley polarization in biased MoS$_2$ bilayers[24].

**Acknowledgments:**

This work was supported by the Research Grant Council of Hong Kong (HKU 706412P).

**Tables**

|  | $a$(Å) | $\Delta$ | $t$ | $2t_\perp$ | $2\lambda$ | $\cos 2\alpha$ | $\langle\sigma_z\rangle_{K\downarrow}$ |
|---|---|---|---|---|---|---|---|
| MoS$_2$ | 3.160 | 1.766 | 1.137 | 0.086 | 0.147 | 0.863 | 0.863 |
| WS$_2$ | 3.153 | 1.961 | 1.436 | 0.109 | 0.421 | 0.968 | 0.962 |
| MoSe$_2$ | 3.288 | 1.541 | 0.951 | 0.106 | 0.182 | 0.864 | 0.855 |
| WSe$_2$ | 3.280 | 1.698 | 1.233 | 0.134 | 0.456 | 0.959 | 0.944 |

**Table 1. Fitting result from *ab initio* band structure calculations.** $\Delta$, $t$, and $\lambda$ (in units of eV) are fitted from the monolayer band structures. Interlayer hopping $t_\perp$ is read out from the valence band splitting at *K* points of bilayers in the absence of spin-orbit coupling. The structural constants use the bulk values. The column $\cos 2\alpha$ is evaluated using Eq. (4) with the listed values of $t_\perp$ and $\lambda$, and according to our band model it gives the layer polarization of the Bloch state $|K\downarrow\rangle$. The last column is the layer polarization directly evaluated from the *ab initio* wavefunction of $|K\downarrow\rangle$. See also supplementary information.

**Figures**

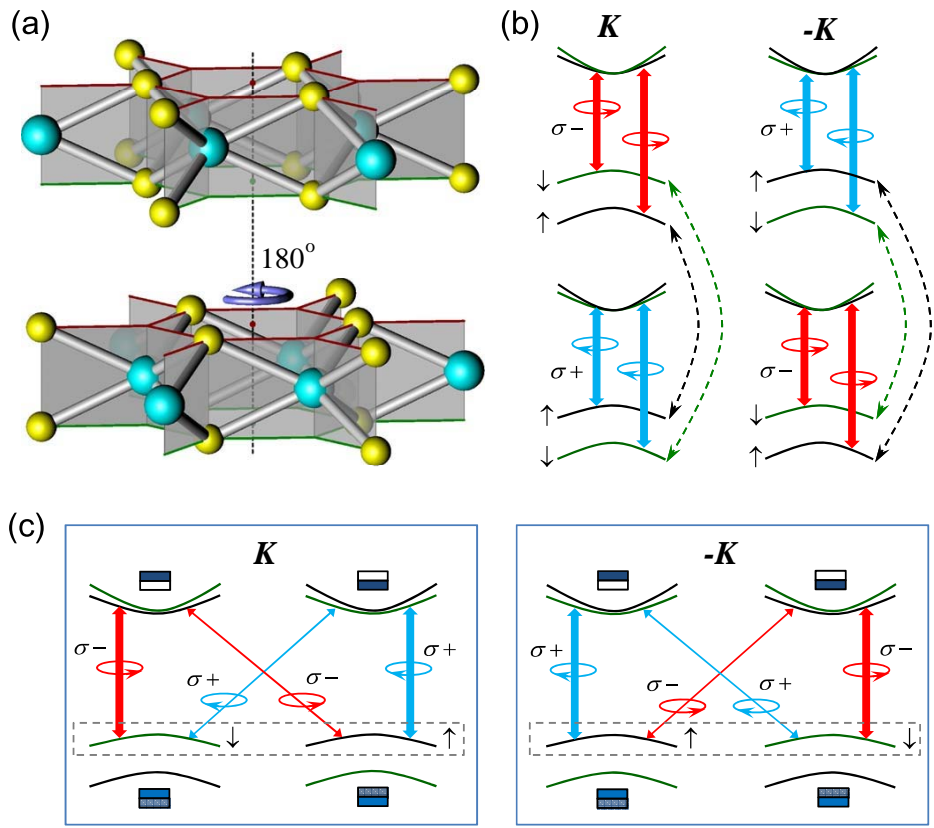

**Figure 1. Group-VI TMDC bilayer. a,** Coordination structure of AB-stacked bilayer. **b,** Optical transition selection rules in decoupled monolayers. Dashed arrows indicate interlayer hopping. **c,** Bilayer optical transition selection rules in the $K$ and $-K$ valleys in the presence of interlayer hopping. Thickness of the arrows represents the transition strength. The layer polarization of the Bloch states is schematically illustrated with the rectangular blocks where darker color denotes more occupation.

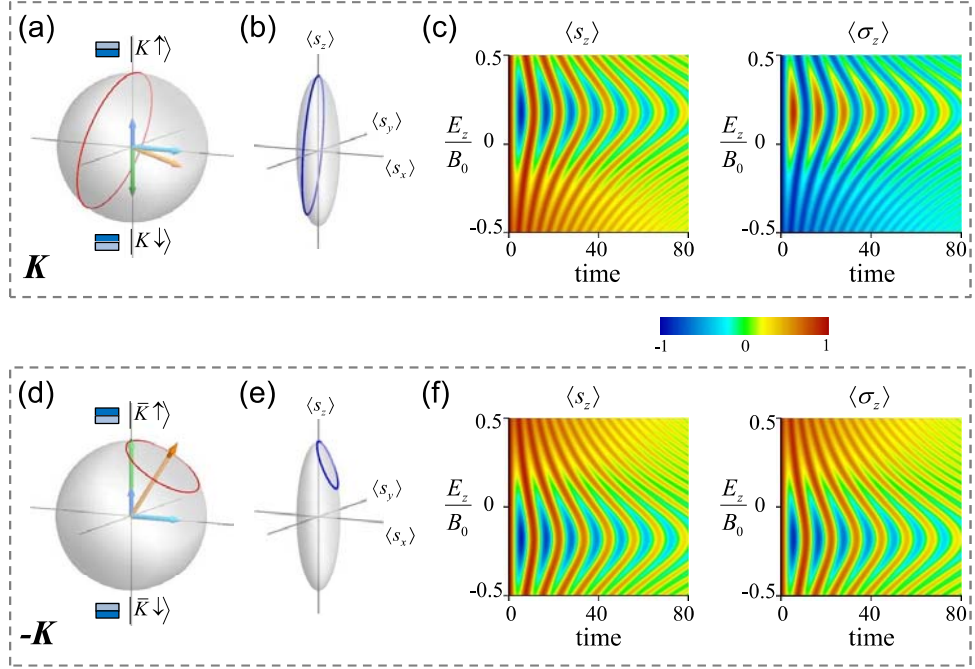

**Figure 2. Valley dependent oscillations of spin and layer polarizations. a,** Bloch sphere spanned by superposition states of the doublet $\{|K\uparrow\rangle, |K\downarrow\rangle\}$. Dark and light blue arrows denote the effective fields from the magnetic field components $B_z$ and $B_x$ respectively, and green arrow denotes that from the electric field $E_z$. Yellow arrow is the total effective field. **b,** Anomalous precessions of the spin vector $\langle s \rangle$ when layer is traced out. **c,** Oscillations of the spin ($\langle s_z \rangle$) and layer ($\langle \sigma_z \rangle$) polarizations in magnetic field $(B_x, B_z) = (B_0 \cos 10°, B_0 \sin 10°)$, from the initial state $|K\uparrow\rangle$. The oscillations are electrically tunable. **d-f,** Same plots in valley -K. The unit of time is $1/B_0$.

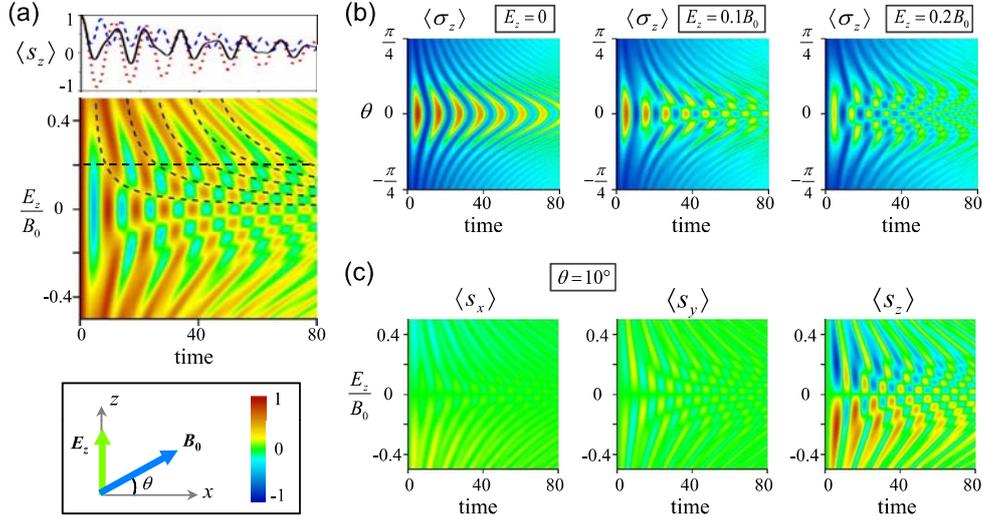

**Figure 3. Magnetoelectric effects. a,** Spin precession in magnetic field along $\theta = 10°$ (c.f. inset), from initial state $\rho_s = 0.5|K \uparrow\rangle\langle K \uparrow| + 0.5|\bar{K} \uparrow\rangle\langle \bar{K} \uparrow|$. Oscillation in $\langle s_z \rangle$ is tunable by the electric field $E_z$ as shown by the contour plot. The dashed curves on the contour image denote the time when the spin precessions at $K$ and $-K$ acquire a phase difference of $(2n+1)\pi$. A horizontal cut of the contour image at $E_z = 0.2B_0$ is shown as the black solid curve atop, while the red and blue curves denote $\langle s_z \rangle$ at $K$ and $-K$ respectively. **b,** Oscillation of the layer polarization $\langle \sigma_z \rangle$ in a magnetic field, with a pattern tunable by $E_z$. The initial state is $\rho_E = 0.5|K \uparrow\rangle\langle K \uparrow| + 0.5|\bar{K} \downarrow\rangle\langle \bar{K} \downarrow|$. **c,** Finite spin polarization can evolve out of the initial state $\rho_E$ of zero spin polarization. In all calculations, we assumed the doublet has a spin relaxation rate of $0.01B_0$. The unit of time is $1/B_0$.

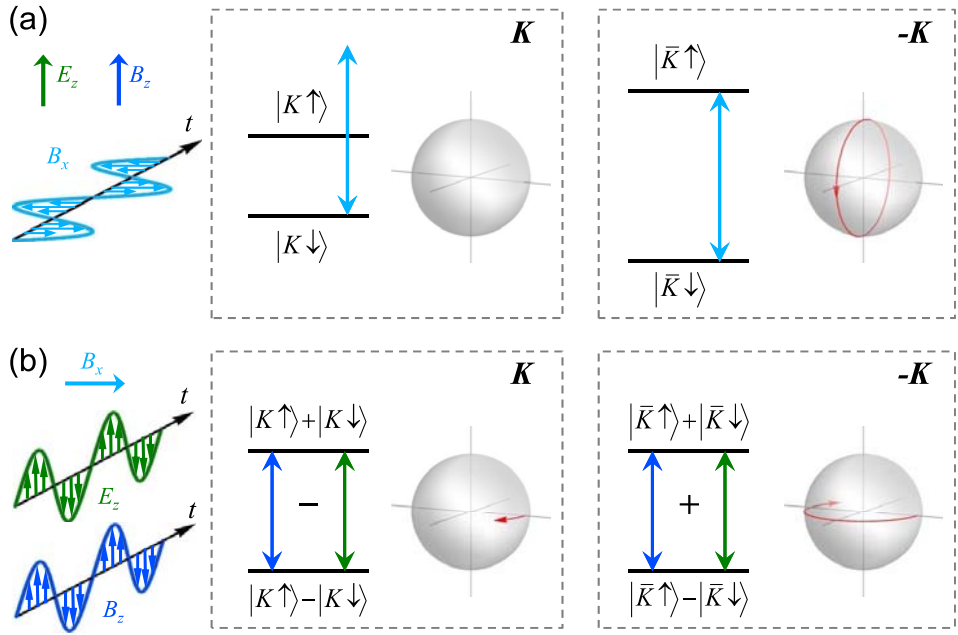

**Figure 4. Valley controlled quantum gates of a spin qubit. a,** Static electric ($E_z$) and magnetic ($B_z$) fields induce a valley dependent splitting of the spin qubit. A pulse of oscillating magnetic field $B_x$ can rotate the spin qubit conditional upon the valley state. **b,** When a static $B_x$ splits the spin qubit, both oscillating $E_z$ and $B_z$ can drive the spin resonance, and the two fields constructively (destructively) interfere in valley -*K* (*K*), which realize valley controlled rotation of the spin qubit about the *z*-axis.